\documentclass[]{aa} % for the letters 
\usepackage{graphicx}
\usepackage{txfonts}
\begin{document}
   \title{Exploring the nature of new main-belt comets with the 10.4m GTC telescope: (300163) 2006 VW139}

%   \subtitle{I. Overviewing the $\kappa$-mechanism}

   \author{J. Licandro\inst{1,2}
          \and
          F. Moreno\inst{3}
          \and
          J. de Le\'on\inst{4}
          \and
          G. P. Tozzi\inst{5}
         \and
          L. M. Lara\inst{3}
          \and 
          A. Cabrera-Lavers\inst{1,2,6}
          }

   \institute{Instituto de Astrof\'isica de Canarias (IAC),
             C/V\'ia L\'actea s/n, 38205 La Laguna, Tenerife, Spain.\\
              \email{jlicandr@iac.es}
         \and
             Departamento de Astrof\'{\i}sica, Universidad de La Laguna (ULL), 
	     E-38205 La Laguna, Tenerife, Spain.\\
         \and
             Instituto de Astrof\'isica de Andaluc\'ia - CSIC, Granada, Spain.\\
         \and
             Departmento de Edafolog\'ia y Geolog\'ia, Universidad de La Laguna (ULL), Tenerife, Spain.\\
        \and
             INAF, Osservatorio Astronomico di Arcetri, Italy.  \\
        \and
            GTC Project, E-38205 La Laguna, Tenerife, Spain.
     }

   \date{Received XXX xx, XXX; accepted XXX xx, XXX}

% \abstract{}{}{}{}{} 
% 5 {} token are mandatory
 
  \abstract
  {}
 %  {Main belt comets (MBCs) are object in orbits typical of main belt asteroids that have been observed with a comet-like coma and/or tail. To understand the mechanism of activation and their origin provide information of cosmogonical interest.  }
   {We aim to study the dust ejected by main-belt comet (MBC) (300163) 2006 VW$_{139}$  to obtain information on the ejection mechanism and the spectral properties of the object, to see if they are compatible with those of ``normal" comets. }
   {Broad-band images in the $g$ and $r$ band as well as a low-resolution spectrum in the 0.35 -- 0.9 $\mu$m region were obtained with the GTC telescope (Roque de los Muchachos Observatory, La Palma, Spain). Images were analyzed to produce a color map and derive a lower limit of the absolute magnitude $H$. A Monte Carlo scattering model was used to derive dust properties such as mass loss rates and ejection velocities as a function of time. The reflectance spectrum  was compared to that of the well-studied MBC 133P/Elst-Pizarro. The spectrum was also used to search for  $CN$ emission and to determine the upper limit of the $CN$ production rate.  }
   {The reflectance spectrum of 2006 VW$_{139}$  is typical of a C-class asteroid, with a spectral slope $S'_V$=0.5$\pm$1.0\%/1000\AA. It is  similar to the spectrum of 133P and other MBCs. No CN emission is detected in the spectrum. A CN production rate upper limit of $3.76 \times 10^{23}$ $s^{-1}$ is derived.  Images show that the MBC present a narrow almost linear tail that extends up to 40.000 km in the anti-solar direction and more than 80.000 km in the direction of the object's orbital plane. The color of the tail is slightly redder than the Sun ($S'$ between 3-6 \%/1000\AA). The Monte Carlo dust tail model derived the mass loss rates and ejection velocity as a function of time, and the results show that the activity onset occurs shortly after perihelion, and lasts about 100 days; the total ejected mass  is $\sim 2\times$10$^6$ kg.}
   {The spectrum of  2006 VW$_{139}$ suggests that it is not a ``normal'' comet. The spectrum is typical of the other observed MBCs. Even if no $CN$ emission is detected, the more likely activation mechanism is water-ice sublimation. Like other well studied MBCs, (300163) 2006 VW$_{139}$ is likely a primitive C-class asteroid that has a water-ice subsurface depth reservoir that has recently  been exposed to sunlight or to temperatures that produce enough heat to sublime the ice. }

   \keywords{minor planets, asteroids: individual: (300163) 2006 VW139 -- comets: general 
-- techniques: photometric -- techniques: spectroscopic
               }

   \maketitle
%
%________________________________________________________________

\section{Introduction}

Asteroid (300163) 2006 VW$_{139}$ (hereafter  VW139) is the eighth main-belt asteroid observed to be ``active", i.e., with a cometary-like dust tail or coma (Hsieh et al. \cite{Hsiehetal2011}). With a semi-major axisof $a$ = 3.0516 AU, an eccentricity $e$=0.2011, and inclination $i$=3.2392$\,^{\circ}$, VW139 fits into the category of  main-belt comets (MBCs) following the pure phenomenological definition by Hsieh \& Jewitt (\cite{Hsieh06}).

Images and spectroscopy of VW139 have been presented in Hsieh et al. (\cite{Hsiehetal2012}). The authors reported that the object presents a tail aligned with the orbital plane of about 1 arcmin, and a short anti-tail  of $\sim$ 10". They claimed that this implies that the observed dust was produced by the action of a long-lived, sublimation-driven emission mechanism. They also reported that there is no evidence of any CN emission in the object spectrum and established an upper limit for the CN production rate of   $Q_{CN} <$ $1.3 \times 10^{24}$ $s^{-1}$. 

Dynamical simulations show that MBCs are extremely unlikely to originate in the place where comets came from: the trans-Neptunian belt (TNB) and the Oort Cloud (e.g., Fern\'andez et al. \cite{Fernandez2002}). They likely formed in situ. This is the case  for VW139 (Hsieh et al. \cite{Hsiehetal2012}), which could also be a member of an asteroid collisional family (Novakovic et al. \cite{Novakovic2012}). 
A recent work suggests that some icy trans-Neptunian objects (TNOs) might have been delivered to the asteroid belt during the Late Heavy Bombardment (Levison et al. \cite{levison2009}), but even those simulations fail to produce low-inclination, low-eccentricity orbits such as those of MBCs. 

%Comets are known to originate in the trans-neptunian belt (TNB) and the Oort Cloud. Observationally, comets  are distinguished from asteroids by the presence of a coma and/or tail.  However, this is not a conclusive criterion. e.g. icy objects only develop a coma if the temperature is sufficient to sublimate ices, so distant comets may not show activity.

%In this paper we define that an object has a cometary origin if it shows similar physical properties to objects that have been considered comets up to now, i.e.,  objects scattered from the trans-neptunian belt or the Oort cloud. 

%The Tisserand parameter with respect to Jupiter ($T_J$) provides a simple way to discriminate dynamically between asteroids and comets (Kresak \cite{Kresak82}; Kosai \cite{Kosai92}). Main-belt asteroids move in orbits with $T_J$ $>$ 3, while comets have unstable orbits with $T_J$ $<$ 3.

%Recently, some objects in orbits indistinguishable  from those of other main belt asteroids have been observed ``active" (with a dust coma and/or tail). They have been called Main Belt Comets (MBCs) (Hsieh \& Jewitt \cite{Hsieh06}).  133P/Elst-Pizarro (hereafter, 133P) and 176P/LINEAR (hereafter, 176P) are also identified as asteroids 7968 and 118401, respectively. They have $T_J$ = 3.184 and 3.166, respectively, and they are two of the seven objects observed so far that can be considered as MBCs following Hsieh \& Jewitt (\cite{Hsieh06}) pure phenomenogical definition. 

The asteroid 133P/Elst-Pizarro is the first discovered MBC (Elst et al. \cite{Elst96}) and so far the best characterized.  It has been observed to be active at every perihelion passage around the orbital quadrant following perihelion, and to be inactive in the two quadrants around aphelion (Boehnardt et al. \cite {Boehnhardt1997}; Toth \cite{Toth06}; Hsieh et al. \cite{Hsieh04, Hsieh09, Hsieh10}; Bagnulo et al. \cite{Bagnulo2010}). This recurring activity supports the hypothesis that 133P has an ice reservoir, which periodically sublimates and elevates dusty material into the coma and tail region. The existence of ice in MBCs is surprising, as they are close to the Sun, but recent thermal models suggest that water ice can survive at greater subsurface depth, even in the asteroid belt (Schorghofer, \cite{Schorghofer}).

The MBCs 133P and 176P/LINEAR  are members of the Themis collisional family.  These two MBCs share surface properties with other Themis family asteroids (see Licandro et al. \cite{Licandroetal2007,Licandroetal2011b}) which indicates that they have formed in situ.  Water ice has recently been discovered on the surface of (24) Themis (Campins et al. \cite{campins2010}; Rivkin \& Emery \cite{rivkin2010}) and (65) Cybele (Licandro et al., \cite{Licandroetal2011a}).
Water ice on the surface of (24) Themis likely  comes from the interior, which supports the hypothesis that there is a water ice reservoir below the surface of some MBCs. The spectral properties of 133P and 176P are significantly different from those of ``normal'' comets,  i.e., comets scattered from the Oort Cloud or  the TNB,  which argues against a cometary origin (Licandro et al. \cite{Licandroetal2011b}).   

Hsieh et al. (\cite{Hsiehetal2012}) also presented numerical simulations showing that the asteroid is dynamically stable for $>$ 100 Myr and searched for a potential asteroid family around the object.  ``A cluster of 24 asteroids within a cutoff distance of 68 $m s^{-1}$. At 70 $m s^{-1}$, this cluster merges with the Themis family, suggesting that it could be similar to the Beagle family".

%Licandro et al. (2007b) suggested also that other near earth asteroid-comet transition objects like (3200) Phaethon and (4015) Wilson-Harrington are scattered MBCs. Dynamical studies indicate it is highly unlikely they have been captured from other sources (Haghighipour 2008). They must have formed in the Main Belt region, presumably at a time when the Snow Line was at $\sim$3AU, during the first Myr of the Solar System. Thermal models suggest that water ice can survive at greater subsurface depth, even in the asteroid belt (Schorghofer 2008). 
%The dust coma and tail, and not a gas emission feature, are so far the only signatures of ongoing activity in MBCs. 
%On the other hand, 176P is the third known MBC and it exhibited cometary activity only at the time of its discovery in 2005 (Hsieh et al. \cite{Hsieh06}). It was never reported active since.

Understanding the origin of MBCs is crucial. If they are formed in situ and if their activity is caused by water ice sublimation, there should be water ice in many asteroids and they must have formed in the main-belt region, presumably at a time when the snow line was at $\sim$3 AU, during the first Myr of the solar system. If they are captured TNBs or Oort Cloud comets, the mechanisms that drove them to their present orbits need to be understood.
The existence of ice in main-belt objects is surprising given their proximity to the Sun, and presents intriguing opportunities for constraining the temperature, composition, and structure of primitive asteroids and our protoplanetary disk.  It is therefore critical for understanding the physical conditions and the mechanisms of planetary formation, and it also addresses the question of the origin of Earth's water: if the outer main belt has a large population of asteroids with ice, they could have contributed to the water on Earth. Finally, this indicates the extent and origin of volatiles in asteroids that could be used as resources for space exploration. 

Late  in 2010 we started a target-of-opportunity program at the 10.4m Gran Telescopio Canarias (GTC), located at the El Roque de los Muchachos Observatory (La Palma, Canary Islands, Spain) to obtain images and, when possible, visible spectra of new discovered MBCs. VW139 is the second MBC observed within the frame of this program  (see Moreno et al. \cite{Moreno2010}), and in this paper we present the resulting data. Images and spectra together with details on data acquisition and reduction are described in Section 2. An analysis of the data is presented in Section 3. In Section 4 a Monte Carlo scattering dust model is applied to fit the observed dust tail and derive the dust properties, and to obtain information on the activation mechanism. Finally, the conclusions are presented in Section 5.

\section{Observations and Data Reduction}

Broad-band images and low-resolution spectroscopy of VW139 were made in service mode on November 29, 2011 using the Optical System for Imaging and Low Resolution Integrated Spectroscopy (OSIRIS) camera-spectrograph (Cepa et al., \cite{Cepa00}; Cepa, \cite{Cepa10}) at the 10.4m Gran Telescopio  Canarias (GTC), located at the El Roque de los Muchachos Observatory (La Palma, Canary Islands, Spain). The OSIRIS instrument consists of a mosaic of two Marconi CCD detectors, each with 2048x4096 pixels and a total unvignetted field of view of 7.8'x7.8', giving a plate scale of 0.127 ''/pix. To increase the signal-to-noise ratio (S/N) for our observations, the data were 2x2 binned, corresponding with the standard operation mode of the instrument. 

The observing geometry is presented in Table\ref{Table1}.

\begin{table}
\centering
\begin{tabular}{l c c c c c}
\hline
\hline
Date (UT) & $r$ (AU) & $\Delta$ (AU)  & $\alpha$($^{\circ}$) & $Sun_{PA}$($^{\circ}$) & $Motion_{PA}$($^{\circ}$) \\ \hline
29/11/2011 & 2.52 & 1.67 & 14.0 & 65.2 & 247.5 \\  \hline
\end{tabular}
  \caption{Observing geometry. $r$ and $\Delta$ are the heliocentric and geocentric distance, $\alpha$ is the phase angle, $Sun_{PA}$ and $Motion_{PA}$ are the position angles of the extended Sun--target radius vector and the negative of the target's heliocentric velocity vector.}
  \label{Table1}
\end{table}

\subsection{Imaging}

CCD images were collected under photometric conditions, dark moon, and with an average seeing of 0.9". Four consecutive series of 5x30 secs in the Sloan $g$ and $r$ filters were obtained between 21:05 and 22:29 UT. Bias correction, flat-fielding, and bad-pixel-masking were performed using standard procedures. The images were finally aligned on the MBC. 

Images were flux-calibrated using Sloan photometric standards. Each group of five ``on-object" aligned images were finally median-averaged to obtain four combined images in $g$  band and another four images in $r$ band. In the first two $g, r$ series the asteroid  was too close to a very bright star, therefore only the images of the last two series were used for the analysis. 

\begin{figure}
   \centering
     \includegraphics[width=9cm]{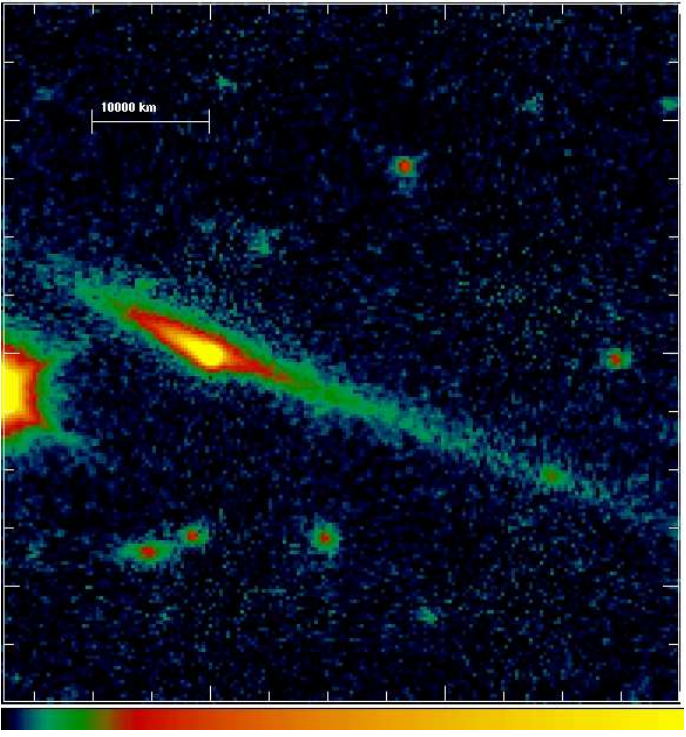}
      \caption{$Af$ calibrated SLOAN $r$-band image of VW139. North is up, ast is to the left.  The antisolar direction is at PA=65.2$\,^{\circ} $. The image covers a region of 75000$\times$75000 km$^2$. This is the last image of the series acquired in $r$, i.e., the one with the near-by bright star at the maximum distance (visible in the E direction). The color look-up table is logarithmic, with black (left side of the color bar) corresponding to $Af$=0 and yellow (right side of the bar) to $Af$=$1\times10^{-7}$.}
         \label{figure1}
   \end{figure}

A constant sky level, measured in image regions out of the observed tail, was subtracted from the images, and they were finally flux-calibrated  in $Af$, i.e., the albedo multiplied by the filling factor of the dust in the coma (A'Hearn et al. \cite{Ahearnetal95}; Tozzi et al. \cite{tozzi2007}), using the following formula

\begin{equation}
%Af=(\frac{2\times2.06\times10^{5}r_h}{dx})^{2}Cs\times10^{-0.4(Zp-Ms)}
Af = 5.34 \times 10^{11} \left(\frac{r_h}{dx}\right)^2 C_s \times 10^{-0.4(Z_p-M_s)},
\end{equation}   

where $r_h$ is the heliocentric distance in AU, $dx$ is the detector pixel size in arcsecs, $C_s$ is the pixel signal in $e^-$/s, and $Z_p$ and $M_s$  are the zero points and the solar magnitude in the used filter.  An example of one of the $r$-band-calibrated images is shown in Fig. \ref{figure1}. 
% with zeropoints of 28.77$\pm$0.01 and 29.27$\pm$0.01 for Sloan $g$ and $r$ bands, respectively (those zeropoints correspond to magnitudes at airmass equal 0, following the instructions on the OSIRIS user manual\footnote{www.gtc.iac.es/en/pages/instrumentation/osiris.php}). 

%=== Figura 1 ===

\subsection{Spectroscopy}
  
\begin{figure}
   \centering
     \includegraphics[width=9cm]{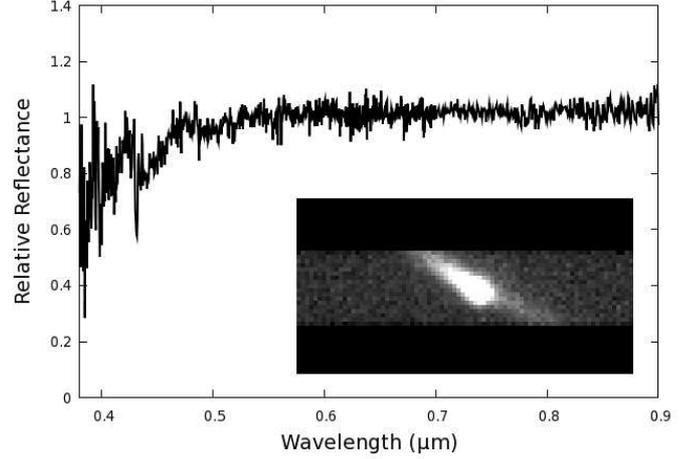}
      \caption{ Final reflectance spectrum of VW139. We also show an image of the asteroid inside the slit.}
         \label{figureSP}
   \end{figure}

The visible spectrum of VW139  was obtained on the same night immediately after the images. We used the R300B and R300R grisms, with dispersions of 2.48 and 3.87 \AA/pixel, respectively. A 5'' slit width was used, oriented toward the parallactic angle to minimize slit losses due to atmospheric dispersion. An image of the object in the slit is shown in Fig.\ref{figureSP}.  Three spectra with an exposure time of 300 sec each were obtained with each grism, covering the 0.35--0.70 $\mu$m and 0.50--1.0 $\mu$m spectral ranges at an airmass of $X= 1.1$. 
%The object was shifted in the slit direction by 6'' between consecutive spectra to better correct the fringing. 

Images of the spectra were bias- and flat-field-corrected with lamp flats. The two-dimensional spectra were wavelength-calibrated with Xe+Ne+HgAr lamps. After the wavelength calibration, sky background was subtracted. We fitted a first-order constant sky level using a section across the spatial direction, and selecting two regions close enough to the MBC, but carefully avoiding any contribution from the observed tail. This procedure was repeated across the spectral direction every 50 pixels. After background subtraction, a one-dimensional spectrum of the optocenter of the MBC was extracted using a very narrow extraction aperture of only $\pm$ 3 pixels (the FWHM measured for the central part of the MBC is $\sim$ 6 pixels) to minimize the dust contribution. To correct for telluric absorption and to obtain the relative reflectance, a G star from the Landonlt  (\cite{landolt92}) list was observed using the same spectral configuration at an airmass ($X=1.2$) similar to that of the object, and was used as a solar analogue star. Each individual spectrum of the object was then divided by the corresponding spectra of the solar analog and then normalized to unity at 0.55 $\mu$m. The three individual spectra were averaged, obtaining a mean spectrum for each of the two grisms. Those spectra were finally merged using the common wavelength interval 0.60--0.70 $\mu$m. The final reflectance spectrum in the 0.38--0.9 $\mu$m region is shown in Fig. \ref{figureSP}. 
%{\bf Out of this spectral region the spectrum was too noisy} {\it JAVIER, NO SE SI QUITAR ESTA ULTIMA FRASE, EN REALIDAD NO APORTA GRAN COSA, NO CREES?}. 

The MBC spectrum was also  flux-calibrated using observations of the standard star G191 (observed the same night) and standard procedures.
The flux-calibrated spectrum of VW139  was used to derive an upper limit for the CN production rate (see Sect. 3.2). 

%=== Figura 2 ===

 \section{Data analysis}
 
 \subsection{Analysis of the images}
 
%&In order to better study the physical properties of VW139, we represent Fig. \ref{figure1} using polar coordinates. The resulting image can be seen in Fig. \ref{figurePolar} . 

In Fig. \ref{figure1} the object clearly presents a narrow  and almost linear tail that extends up to 40.000 km in the anti-solar direction (position angle PA$\sim$ 63$\,^{\circ} $) and more than 80.000 km in the direction of the object's orbit plane (PA$\sim$ 250$\,^{\circ} $).  This image is typical of comets when observed with the observer very close to the  comet's orbital plane and, as we show in Sect. 4, implies a dust ejection that lasted several months.

%We first studied possible color variations within the tail by dividing the $r$-band by the $g$-band images and deriving the spectral slope $S'$ (see Fig. \ref{figureColor}). Notice that the tail in the anti-solar direction is significantly redder. This is an indication that the larger particles concentrate in this part of the tail. The maximum spectral slope of the tail is $\simeq$ 5\%/1000\AA ~close to the nucleus in the solar direction (the red color) and the minimum one is $\simeq$ -15\%/1000\AA ~in the opposite direction (the green color). 

% VERSION VIEJA We first studied possible color variations within the tail by dividing the $r$-band by the $g$-band  images (see Fig. \ref{figureColor}). Close to the nucleus in the solar direction (the ``anti-tail''), the $r$/$g$ pixel values are larger (red color) than in the tail direction (in green), which means that the anti-tail is significantly redder than the tail. This suggests that the larger particles concentrate in the anti-tail direction.  

% NEW VERSION
%We first studied possible color variations within the tail by computing the spectral slope $S'$  from  the $r$-band by the $g$-band  images (see Fig. \ref{figureColor}) using 

%{\bf We first studied the color  of the dust tail} by computing the spectral slope $S'$  from  the $r$-band by the $g$-band  images (see Fig. \ref{figureColor}) using 
We first studied the color  of the dust tail by computing the spectral slope $S'$  from  the $r$-band by the $g$-band  images using 

\begin{equation}
%S' = (F_r/F_{rS}-F_g/F_{gS}) * F_{gS}/F_g) / (6.228 - 4.844),
S' = (F_r/F_g * F_{gS}/F_{rS} - 1) / (6.228 - 4.844),
\end{equation}   

where $F_r$ and $F_g$ are the flux-calibrated $r$- and $g$ images, and $F_{gS}$ and $F_{rS}$ are the flux of the Sun in the $g$ and $r$ band. The $F_{gS}/F_{rS} $ ratio is computed by using the solar color $(g-r)_S$=0.44. The central wavelength of the $g$ and $r$ filters are 4844 and 6228 \AA.
%computed by using the solar color $(g-r)_S$=0.44 (see http://www.astro.umd.edu/~ssm/ASTR620/mags.html). The central wavelength of the $g$ and $r$ filters are 4844 

%Close to the nucleus in the solar direction (the ``anti-tail''), $S' \sim 6$\%/1000 \AA ~  (red color in Fig. \ref{figureColor}), a slightly redder slope than in the tail direction ($S' \sim 4-3$\%/1000 \AA in green in the figure). This suggests that the larger particles concentrate in the anti-tail direction.  

%{\bf The color of the dust in the tail is slightly redder than the Sun, with $S' $ between $\sim 6$\%/1000 \AA ~  (red color in Fig. \ref{figureColor}) and 4-3\%/1000 \AA ~(in green in the figure).  }
The color of the dust in the tail is slightly redder than the Sun, with $S' $ between 6\% and 3\%/1000 \AA .

%==

The absolute magnitude of VW139 is $H$=16.1 according to the Minor Planet Center  (MPC), and was computed when the asteroid was not active. We performed aperture photometry of the MBC using an aperture radius of  5" to determine a lower limit for the asteroid magnitude (asteroid + ejected dust in the aperture) and a lower limit for $H$ on both filters. 

We applied this only to the last two combined images of each filter (because a bright star was too close to the asteroid in the first two images). The measured magnitudes are $r$=19.07 and 19.14, and $g$= 19.60 and 19.63, and thus the corresponding absolute magnitude lower limits are $H_r$= 15.16$\pm$0.04 and $H_g$=15.67$\pm$0.02 (assuming the typical $G$ value of 0.15 of the IAU H,G system). Of course this is a lower limit, because the dust of the tail contributes to the asteroid luminosity. The results are compatible with the value of $H$ given by the MPC. Considering that we assumed $G=0.15$, the uncertainties on the derived $H_r$ and $H_g$ lower limits are about 0.1-0.2 mag. Our $H_r$ determination is compatible with the $H_r = 15.1--15.0\pm0.1$ reported by Hsieh et al. (\cite{Hsiehetal2012}) between Nov. 05 and Dec. 04, 2011.

\subsection{Analysis of the spectrum}\label{specanalysis}

The obtained reflectance spectrum of the optocenter of VW139 is very similar to that of a C-type asteroid, almost neutral in the 0.5--0.9 $\mu$m region (with a spectral slope $S'_V$=0.5$\pm$1.0\%/1000 \AA), and a UV drop below 0.55 $\mu$m.  

Because the object was active, the obtained spectrum is ``contaminated" by scattered light from the dust tail.  The photometry reported in the previous section and the absolute magnitude reported in the MPC show that in an aperture of 5 arcsec, the flux from the dust is about twice that reflected by the asteroid surface. Hsieh et al. (\cite{Hsiehetal2012}) reported a value of 2.5. For that reason, and considering that the PSF of the asteroid is much steeper than that of the tail,  we used  a very small extraction aperture of 0.8 arcsec to minimize the dust contribution.  In this small aperture it is too difficult to evaluate how significant the contribution of the dust is, but in any case it is less than for a larger aperture.  The  difference in color between this spectrum ($S'_V$=0.5) and the color measured in the tail in the previous section ($S'$=0.5) suggests that a large part of the light in this aperture comes from the asteroid surface.

However, the color of the dust close to the asteroid is slightly redder than the Sun, as shown in the previous section. The spectra of the dust cometary coma are usually featureless and reddish in the wavelength range of the presented VW139 spectrum, but  the spectrum of the asteroid itself  might be bluish, as is typical of a B-type asteroid.  A spectrum of VW139 when it is inactive is desiderable to accurately determine its taxonomical classification.

A bluish to almost neutral  spectral slope is atypical of active comets and in cometary nuclei (Kolokolova et al. \cite{kolokolova2004}; Licandro et al., \cite{Licandroetal2007}). 

\begin{figure}
   \centering
      \includegraphics[width=9cm]{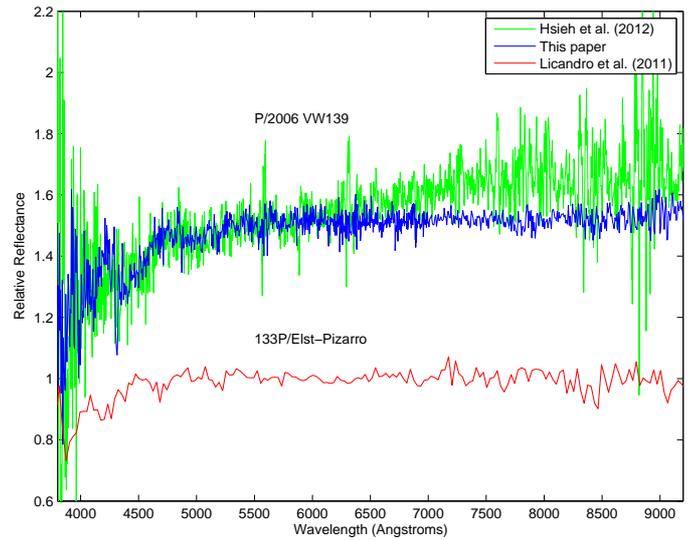}
      \caption{Comparison of the reflectance spectra of  VW139 presented in this paper (blue) and  that from Hsieh et al. (2012) (green).  We have also included in the plot the reflectance spectrum of 133P/Elst-Pizarro  (red) from Licandro et al. (2011b). The two spectra of VW139 are shifted  vertically by 0.5 for clarity. 
              }
         \label{figureMBCEP}
   \end{figure}

Hsieh et al. (\cite{Hsiehetal2012}) also published a spectrum of VW139 obtained a week before  our observations. To compare ours with their results, both spectra are shown in Fig. \ref{figureMBCEP}. The spectrum of Hsieh et al. (\cite{Hsiehetal2012}),  with a lower S/N, is very similar to ours below 6500 \AA \ and slightly redder at longer wavelengths.  Hsieh et al. (\cite{Hsiehetal2012}) reported a spectral slope $S'_V$=7.2\%/1000 \AA, computed using the 0.4--0.9 $\mu$m spectral range. If, as is usually done by convention, we only consider the 0.5--0.9 $\mu$m region (to  avoid the UV drop-off), the spectral slope is $S'_V$=5.4$\pm$1.0\%/1000 \AA, still slightly redder than ours. 

The similarities  between both spectra in the blue region  (below 5500 \AA) is a clear indication that the differences are not caused byo problems related to differential extinction (the Hsieh et al.  spectrum was obtained without the orientation toward the parallactic angle) and/or by slit losses, because, in that case, the blue region of the spectrum is the most affected. One possibility is that because the Hsieh et al. spectrum was obtained with the slit aligned  with the tail, maybe the aperture used to extract their spectrum was wider than the one we used (1.5"), and consequently they  might have included a larger amount of dust. 
%Notice that in Sect. 3.1 we showed that the tail in the Sun direction is redder than the rest of the MBC, and therefore including more dust could explain the small difference in color observed.

% === Figura 5 ===

Licandro et al. (\cite{Licandroetal2011b}) analyzed the nature of MBCs based on their spectra. In Fig. \ref{figureMBCEP}  our spectrum of VW139 is also compared to  the spectrum of 133P/Elst-Pizarro published by Licandro et al. The spectra are very similar, therefore the conclusions about the nature of 133P in Licandro et al. (\cite{Licandroetal2011b}) can be reproduced here for VW139: the spectrum of VW139 resembles that of asteroids of the C-complex and  is significantly different from those of the large majority of observed cometary nuclei,  typically D- or P-types. 

We also searched for any signature of gas emission that is typically observed in cometary comae, of which that of CN  (B$^2 \Sigma^+$--X$^2 \Sigma^+;\Delta v = 0$) at $\sim$3800 \AA ~is the most prominent. The detection of  gas emission would be  a strong test to determine if the observed tail and coma are produced by ice sublimation as in normal comets. 

\begin{figure}
   \centering
      \includegraphics[width=6cm,angle=270]{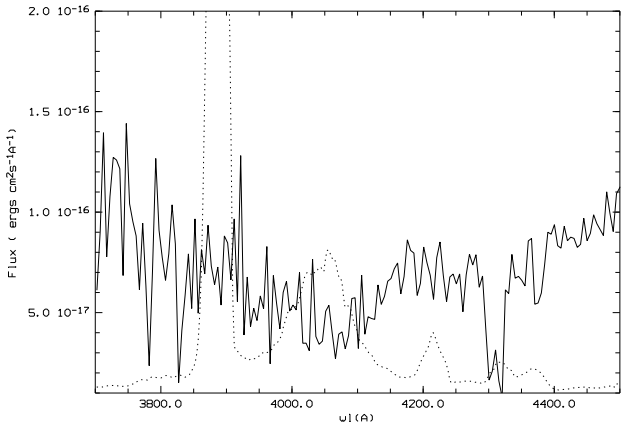}
      \caption{ Visible spectrum of VW139 (solid line) in the region of the CN and C$_3$ emission bands (in ergs cm$^{-2}$ s$^{-1}$ \AA$^{-1}$) compared to the spectrum of active comet 8P/Tutle (dashed line) obtained  on Jan. 1, 2008 (Lara, L. personal communication). The spectrum of VW139 is averaged over  an aperture of 40 pixels, i.e., 10.16 arcsec, meaning 12300 km at the comet distance, and is multiplied by 50 for a better comparison with the spectrum of 8P/Tuttle.}  
       \label{figureCN}
   \end{figure}

In Fig. \ref{figureCN} we show a zoom of our flux-calibrated spectrum of VW139 in the wavelength region of this CN band and compare it with the spectrum of active Jupiter family comet 8P/Tuttle (Lara, L., personal communication). None of the  molecular bands, in particular the CN emission, which is clearly seen in the spectrum of 8P centered at 3870 \AA\ , are detected in the spectrum of VW139. Hsieh et al. (\cite{Hsiehetal2012}) also failed to detect CN emission in their lower S/N spectrum obtained a week before ours.

%We derived an upper limit of the $CN$ production rate $Q_{(CN)} = 2.05^{22} s^{-1}$ and assuming a $Q_{CN}/Q_{OH}$ and $Q_{OH}/Q_{H_2O}$ of Jupiter family comets like in Hsieh et al. (\cite{Hsiehetal2012}) we derive an upper limit for $Q_{(H2O)} = 7.1^{24} s^{-1}$.

An upper limit to the CN production rate is achieved by studying the flux measure in the 3830-3905 \AA\ range in an aperture of 40 pixels, i.e., 10.16 arcsec, meaning 12300 km at the comet distance. We also measured the continuum that borders the band at 3770-3815 \AA\ and 3910-3970 \AA\,  by approximating the continuum contribution by interpolating the left- and right-hand continua. Any remaining flux in the 3830-3905 \AA\ region represents a $1\sigma$ upper limit to the CN emission. 

We converted the emission band  flux into column density using the g-factor from Schleicher (\cite{sch10}) at the asteroid heliocentric velocity of 2.124 km/s and distance of 2.524 AU. The resulting g-factor is g=$3.217 \times 10^{-1}$ erg s$^{-1}$ molecule$^{-1}$. 

To compute the gas production rate, we assumed the Haser modeling (Haser, \cite{haser57}) with the CN parent velocity $v_p$ scaled with $r_h$ ($v_p=0.86 r_h^{-0.4}$ km/s), and $v_d=1$ km/s for CN itself. The CN parent scale length is 13.000 km, whereas the CN one is 210.000 km (A'Hearn et al. \cite{Ahearnetal95}). With this set of parameters, the Haser modeling provides a CN  $1\sigma$ production rate upper limit of $3.76 \times 10^{23}$ $s^{-1}$,  i.e. $3\sigma$ of $1.13 \times 10^{24}$ $s^{-1}$.  

Our determination is on the same order of magnitude as  the one reported by Hsieh et al. (\cite{Hsiehetal2012}),  who obtained a value of $1.3 \times 10^{24}$ $s^{-1}$. 
%This value, on the other hand, puts more strict constraints on the water production rate.

However, this does not mean that there is no gas emission and/or that a reasonable upper limit of the production rate of $H_2O$ ($Q_{H_2O}$) can be derived from it. A limit from $Q_{H_2O}$ is usually determined assuming the average ratio $Q_{CN}/Q_{OH}$ and $Q_{OH}/Q_{H_2O}$ of Jupiter-family comets (e.g. Hsieh et al. \cite{Hsiehetal2012}; Licandro et al. \cite{Licandroetal2011b}). But, as noticed by Licandro et al. (\cite{Licandroetal2011b}), this is probably meaningless, because MBCs likely formed closer to the Sun than ``normal" comets. This implies a higher temperature, therefore one would expect a much lower fraction of species that are more volatile than water ice with respect to water ice, i.e., $Q_{CN}/Q_{OH}$ is probably much lower for the MBCs than for ``normal'' comets.

% === Figura 6 ===

%Para calcular cualquier tasa de producci—n de gas, se ha de quitar el continuo subyacente. Haciendo eso, el upper limit de Q(CN) es m‡s estricto que lo que  H. Hsieh calcula, yo obtengo 2.05e22 s??. Para comprobar con lo que ha hecho H. Hsieh (sospechando que Žl no ha quitado ningœn continuo subyacente), lo he hecho "mal" y mi resultado coincide con el suyo: yo obtengo Q(CN) < 1.6e24 s?? y el dice haber obtenido 1.3e24 s?? ...

%O sea, que las tasas de producci—n de H2O son aœn 2 —rdenes de magnitud por debajo de que H. Hsieh dice, o sea, Q(H2O) < 7.1e24 s??

\section{The dust model}

We analyzed the asteroid images by a direct Monte Carlo dust tail model, which is based on previous works on cometary dust tail analysis  (e.g. Moreno \cite{Moreno2009}; Moreno et al. \cite{Moreno2010}), and on the characterization of the dust environment of  comet 67P/Churyumov-Gerasimenko  before Rosetta's arrival in 2014 (the so-called  Granada model, see Fulle et al. \cite{Fulle2010}). 

Briefly, the code computes the trajectory of a large number of particles ejected from a cometary nucleus surface, which are exposed to solar gravity and radiation pressure. The gravity of the asteroid itself is neglected, which is a good approximation for small-sized objects.  

In an ice-sublimation-driven scenario, the particles are  accelerated by gas drag to their  terminal velocities, which are the input  velocities considered in the model. Particle ejection by any other mechanism (e.g., a collision) can also be simulated.  Once ejected, the particles describe a Keplerian trajectory around the Sun, whose orbital elements are computed from the terminal velocity and the ratio of the  force exerted by the solar radiation pressure and the solar gravity $\beta$ (see Fulle \cite{Fulle89}). This parameter can be expressed as $\beta = C_{pr}Q_{pr}/(2\rho r)$, where $C_{pr}$=1.19$\times$ 10$^{-3}$ kg m$^{-2}$, and $\rho$ is the particle density, assumed to be $\rho$=1000 kg m$^{-3}$. For particle radii $r >$ 0.25 $\mu$m, the radiation pressure coefficient is $Q_{pr}\sim$ 1 (Burns et al. \cite{Burns79}). To compare the model results with the observed isophotes, for CPU and memory reasons, we rebinned the original image by 2x2 pixels so that its spatial resolution becomes 310 km pixel$^{-1}$.

For the observation date, we computed the trajectories of a large number of dust particles and calculated their positions on the $(N,M)$ plane. $M$ is the projected Sun-comet radius vector, and $N$ is perpendicular to $M$ in the opposite half-plane with respect to the nucleus velocity vector. Then, we computed their contribution to the tail brightness.

\begin{figure}
   \centering
      \includegraphics[width=7cm,angle=-90]{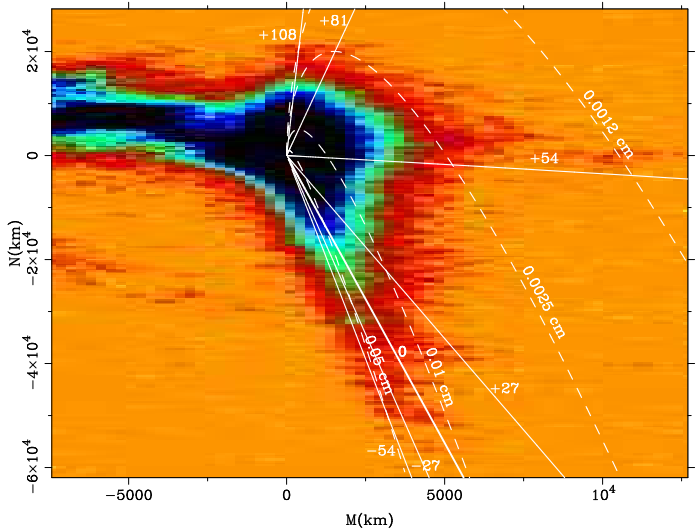}
      \caption{ Syndyne-synchrone diagram overimposed on a VW139 image. The density of the particles is assumed at $\rho$=1000 kg m$^{-3}$. The synchrones are shown as solid white lines, and correspond, in clockwise order, to +108, +81, +54, +27, 0, -27, and -54 days to perihelion, as labeled. Syndynes are shown as white dashed lines, and
correspond to 0.0012, 0.0025, 0.01, and 0.05 cm, as labeled. The $M$- and $N$-axis are scaled independently to avoid label crowding. The bright feature on the left centered at approximate coordinates (-5000,7000) corresponds to a a bright field star that can be seen in Fig. \ref{figure1} placed near the middle of the left ordinate axis.}
         \label{figureModel1}
   \end{figure}

For a first approximation of the input parameters to impose on the dust model, we first obtained a syndyne-synchrone diagram. Figure \ref{figureModel1} shows this map with seven synchrones from +108 to -54 days to perihelion, and three syndynes corresponding to particle radii of 12, 25, and 100 $\mu$m. The $M$- and $N$-axis are scaled independently to avoid label crowding. Based on this map, and since the observation was made about 130 days post-perihelion, we infer that the onset of the asteroid activity should have occurred around perihelion, lasting close to the observation date, as a first approximation. 

The synthetic tail brightness depends on the  ejection velocity law assumed, the particle size distribution, the dust mass loss rate, as well as on the geometric albedo of the particles. We assumed isotropic ejection. To constrain the number of free parameters, the size distribution was assumed to be independent of time and to be characterized by a power-law with index $\alpha$. The maximum size of the ejected particles is set by the  escape velocity. If $H$=16.1, and the albedo is $p_v\sim$0.05, the asteroid radius would be $R\sim$1.8 km. If the bulk density is $\rho$=1000 kg m$^{-3}$, the escape velocity would be about 0.3 m s$^{-1}$, assuming a spherical object. Based on the syndyne-synchrone locus, we  set the lower limit of particle sizes at 1 $\mu$m. Smaller particles ejected some 30 days before the observation would be beyond the image limits. 

\begin{figure}
   \centering
      \includegraphics[width=9cm,angle=-90]{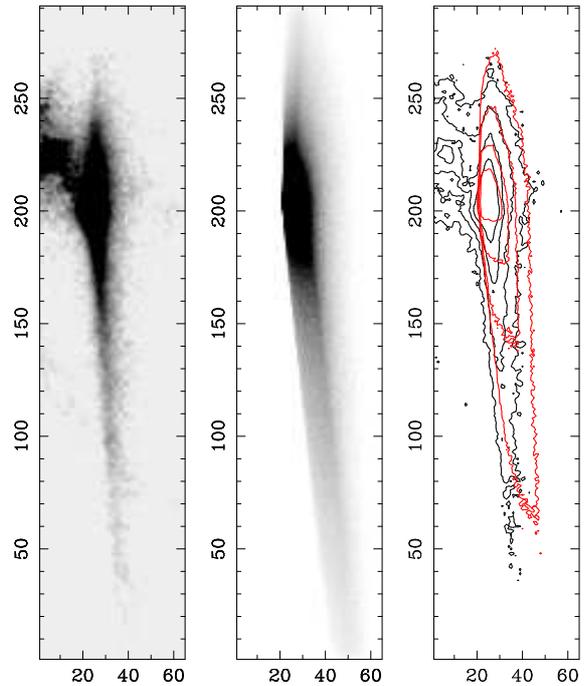}
      \caption{Left: median-averaged image of VW139 obtained on November 29, 2011, referred to the ($N,M$) plane. Center: modeled image. Right: Comparison of observed isophotes (black) and modeled isophotes (red). The contour levels are 5$\times$10$^{-15}$, 10$^{-14}$, and 2$\times$10$^{-14}$, in solar disk intensity units. Each pixel corresponds to 310 km.               }
         \label{figureModel2}
   \end{figure}

\begin{figure}
   \centering
      \includegraphics[width=6cm,angle=-90]{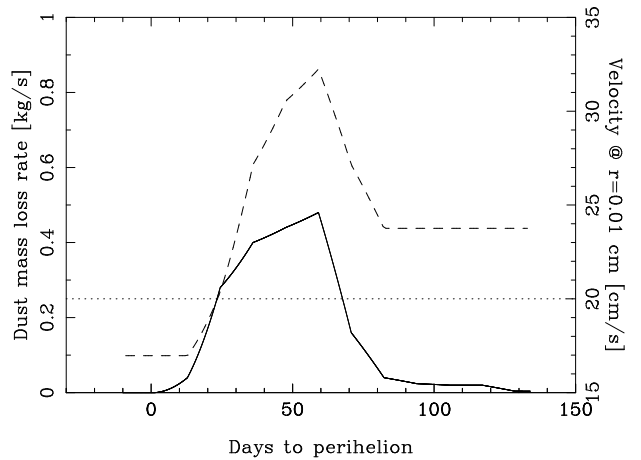}
      \caption{Derived mass loss rate (solid line, left vertical axis) and ejection velocity (dashed line, right vertical axis), as a function of time. The ejection velocity refers to a 0.01 cm particle. The dotted line indicates the assumed escape velocity.}
         \label{figureModel3}
   \end{figure}

Then, we ran the code for a large number of input dust loss masses and particle velocities as functions of the heliocentric distance. The velocity was assumed to vary with $\beta$ as $v \propto \beta^{1/2}$. The best fit is shown in Figure \ref{figureModel2}, whereas the dust loss rates and particle velocities are shown in Figure  \ref{figureModel3}. Our best-fit model used a power-law size distribution function with power index of --3.1, which is in the range of the estimates for other comets and MBCs. For the escape velocity, we found $v_{esc}$=0.2 m s$^{-1}$ as the best-fit. With this value for the escape velocity, we derived from the model that the largest ejected particles forming the anti-sunward branch are about 200 $\mu$m in radius. If the escape velocity is assumed to be higher than 0.2 m s$^{-1}$, the maximum size of those particles would be smaller, and the model would be unable to fit that anti-sunward branch. This is clear from the syndyne-synchrone network of Fig. \ref{figureModel2}, which indicates that this feature can be fitted by particles of size $\sim$100 $\mu$m or larger.  On the other hand, if $v_{esc}$ were slower than $\sim$0.2 m s$^{-1}$, it would be possible to fit the dust feature, because particles in excess of 200 $\mu$m could be ejected.  %However, the implied mass and density would be too low: even for $v_{esc}$=0.2 m s$^{-1}$, the object's density would be already of only 450 kg m$^{-3}$, assuming an spherical object of size $R$=1.8 km. This density estimate should, however, be taken with caution owing to the uncertainties in the object's magnitude and the assumed model parameters.}

%Our best-fit used finally a $v_{esc}$=0.2 m s$^{-1}$, which means that the largest-sized particles were about $r=200 \mu$m.  A higher escape velocity would limit the size to smaller particles, which produce a much worse fit than that shown in Fig. \ref{figureModel2}. This is clear also from the syndyne-synchrone diagram (Fig. \ref{figureModel1}). 

%If $v_{esc}$=0.2 m s$^{-1}$, and the radius of the object is $R$=1.8 km, then its density  would be of only 450 kg m$^{-3}$. Finally, the best-fit power index was --3.1, which is within the range of other estimates for other comets and MBCs. The total mass ejected is 1.9$\times$10$^6$ kg, about 2$\times$10$^{-5}$ \% of the total mass of the asteroid. 

The total dust mass ejected derived from the model is $\sim 2\times$10$^6$ kg, about 2$\times$10$^{-5}$ \% of the total mass of the asteroid. The activity onset is shortly after perihelion, and lasts about 100 days (see Fig. \ref{figureModel3}). This mass loss peaks some 60 days after perihelion at 0.5 kg s$^{-1}$, and  is correlated with the ejection velocity.  This is compatible with the early observations of Hsieh et al. (\cite{Hsiehetal2011}), which describe the object as having a non-stellar PSF on August 30, about 45 days  after perihelion. On the other hand, the velocities are consistent with expectations of cometary activity at 2.5 AU from the Sun, which supports the assumption that water ice sublimation is the mechanism that ejected the observed dust.

% Figura 7 ===

% === Figura 8 ===

% === Figura 9 ===

%\section{Results}
%______________________________________________________________

\section{Discussion and conclusions}

The principal aim of studying MBCs is, as explained in the introduction, to understand their activation mechanism and origin. Are they comets, or asteroids that, under certain circumstances, eject dust? 

Images of the ejected dust can help us to understand what the dust ejection mechanism is. The images of 2006 VW$_{139}$ obtained on November 29, 2011 with the OSIRIS camera-spectrograph at the 10.4m GTC (El Roque de los Muchachos Observatory, La Palma, Canary Islands, Spain) show an object that presents a narrow, almost linear tail that extends up to 40.000 km in the anti-solar direction and more than 80.000 km in the direction of the object's orbit plane. This tail is typical of comets observed very close to the  comet's orbital plane. Moreover, the color map shows that the tail in the anti-solar direction is significantly redder, which indicates that the larger particles concentrate in this  region, as expected for an object that presents a comet-like activity.

Using a direct Monte Carlo dust tail model (e.g. Moreno \cite{Moreno2009}; Moreno et al. \cite{Moreno2010}), we fitted the images and derived the mass loss rates and ejection velocity as a function of time. We first concluded that the observed tail is not compatible with a short-lasting event like a collision or an outburst. The activity onset  occurs shortly after perihelion, and lasts about 100 days. All evidence favors a comet-like ejection mechanism, which means dust ejected by the sublimation of volatiles.

%We were also able to derive other properties of the ejected dust and the object itself. First, the best-fit corresponds with a escape velocity $v_{esc}$=0.2 m s$^{-1}$, so the largest-sized particles were about $r=200 \mu$m, and assuming that the radius of the object is $R$=1.8 km, the bulk density of the asteroid is $\sim$450 kg m$^{-3}$. The very low derived bulk density suggests a highly interior macro-porosity, which is  common in C-type asteroids (Britt et al. \cite{Brittetal2002}). But the derived density is lower than the mean density of the very small number of C-types for which the bulk density is known (Britt et al. \cite{Brittetal2002}). On the other hand, it is similar to the bulk density reported for comet nuclei (Weissman et al. \cite{Weissmanetal2005} ). Anyhow, the conclusions that can be drawn from the bulk density determined should consider that: (1) the derived density strongly depends on the radius $R$, which is very uncertain: if $R$=1.2 km then the mean density will be about 1000 kg m$^{-3}$, compatible with that reported for C-type asteroids; (2) the MBC is considerably smaller than the C-type asteroids with derived density. In any case, VW139 is an object with a low bulk density, something between the density of a comet nucleus  and a C-type asteroid. 

We were also able to derive other properties of the ejected dust and the object itself. From the dust model, and to fit the sunward branch intensities, the best-fit escape velocity derived was $v_{esc}$=0.2 m s$^{-1}$.
%, which implies a bulk density of the asteroid of $\sim$450 kg m$^{-3}$ for an assumed object's radius of $R$=1.8 km. We recognize, however, that this density estimate is not conclusive due to uncertainties in both the object's magnitude and the assumed model parameters. The very low derived bulk density would suggests a highly interior} macro-porosity, which is  common in C-type asteroids (Britt et al. \cite{Brittetal2002}). But the derived density is lower than the mean density of the very small number of C-types for which the bulk density is known (Britt et al. \cite{Brittetal2002}). On the other hand, it is similar to the bulk density reported for comet nuclei (Weissman et al. \cite{Weissmanetal2005} ). Anyhow, the conclusions that can be drawn from the bulk density determined should consider that: (1) the derived density strongly depends on the radius $R$, which is very uncertain: if $R$=1.2 km then the mean density will be about 1000 kg m$^{-3}$, compatible with that reported for C-type asteroids; (2) the MBC is considerably smaller than the C-type asteroids with derived density. In any case, VW139 is an object with a low bulk density, something between the density of a comet nucleus  and a C-type asteroid. 

On the other hand, the mass loss peaks some 60 days after perihelion at 0.5 kg s$^{-1}$, and is correlated with the ejection velocity. This is compatible with the early observations of Hsieh et al. (\cite{Hsiehetal2011}), which describe the object as having a non-stellar PSF (indicative of a dust cloud around the object), on August 30 (45 days after perihelion). The derived velocities are also consistent with expectations of cometary activity at 2.5 AU from the Sun. The total mass ejected is $\sim 2\times$10$^6$ kg.
 
%and low resolution spectroscopy of VW139 were done in service mode on November 29$^{\rm{th}}$, 2011 using the Optical System for Imaging and Low Resolution Integrated Spectroscopy (OSIRIS) .he 

 %The measured magnitudes using a 5" aperture are $r$=19.07 and 19.14, and $g$= 19.60 and 19.63, respectively, thus the corresponding absolute magnitudes lower limits are $H_r$= 15.33$\pm$0.07  and $H_g$=15.84$\pm$0.04. the dust of the tail contribute to the
%asteroid luminosity. The results are compatible with the fainter $H$ of the asteroid alone given by the MPC.

The spectrum of the object can be compared with spectra of comets and primitive asteroids to determine which group the object belongs to, and to search for gas species in the coma (which would confirm of a comet-like activation mechanism). The reflectance spectrum of  the optocenter of VW139 is typical of a C-type asteroid, almost neutral in the 0.5--0.9 $\mu$m region (with a spectral slope $S'_V$=0.5$\pm$1.0\%/1000\AA), and a UV drop below 0.55 $\mu$m.  A bluish to almost neutral slope in the spectrum is atypical of active comets but typical of MBCs (Licandro et al. \cite{Licandroetal2011b}). Accordingly, VW139 presents a spectrum typical of the asteroids with similar semi-major axis and different from those of comet nuclei. This supports an in situ origin, as  for dynamical simulations. 

We finally searched for the CN   (B$^2 \Sigma^+$--X$^2 \Sigma^+;\Delta v = 0$) signature of gas emission at $\sim$3800\AA ~  that is typically observed in cometary comae,  finding no evidence of this feature in our spectrum. We derived an upper limit of the CN production rate $Q_{CN} = 3.76 \times 10^{23}$ $s^{-1}$. Unfortunately, the direct determination of $Q_{H_2O}$ is very difficult, and to derive an upper limit of $Q_{H_2O}$ based on the upper limit of $Q_{CN}$, we had to assume values of the $Q_{CN}/Q_{OH}$ and $Q_{OH}/Q_{H_2O}$ ratios that are typical of comets. This last assumption might not be correct for MBCs, as explained in Sec. \ref{specanalysis}.

In conclusion, (300163) 2006 VW$_{139}$, like other well-studied MBCs (see the case of 133P/Elst-Pizarro) is likely a primitive C-type asteroid that has a water-ice reservoir at subsurface depth.  Owing to still uncertain mechanisms (probably a collision that eroded its surface), this reservoir has recently been exposed to sunlight or to temperatures that produce enough heat to sublime the ice and drag dust from the surface.

\begin{acknowledgements}
We thank H. Hsieh for providing the ascii version of their spectrum and to the anonymous referee who helped to improve this paper.  J.L. acknowledges support from the project AYA2011-29489-C03-02 (MEC). L.M.L's and JdL's work has been funded through project AYA 2009-08011 awarded by the Spanish ``Ministerio de Ciencia e Innovaci\'on''. JdL aslo acknowledges financial support from the current Spanish ``Secretar\'{\i}a de Estado de Investigaci\'on, Desarrollo e Innovaci\'on'' (Juan de la Cierva contract). F.M. was supported by contracts AYA2009-08190 and FQM-4555 (Junta de Andaluc\'{\i}a). Based on observations made with the Gran Telescopio Canarias (GTC), installed in the Spanish Observatorio del Roque de los Muchachos of the Instituto de Astrof\'isica de Canarias, on the island of La Palma.

\end{acknowledgements}

\end{document}